\documentclass[a4paper,12pt]{article}
\usepackage{epsfig}
\date{\today}
\begin{document}

\renewcommand{\thefootnote}{\fnsymbol{footnote}}

%\begin{titlepage}
%\pagestyle{empty}
\rightline{SUSX-TH/02-003}
\vskip 1cm
\begin{center}
{\bf \Large{  Torsion Constraints in the Randall--Sundrum Scenario \\[10mm]}}
{ \large{Oleg Lebedev} \\[6mm]}
\small{Centre for Theoretical Physics, University of Sussex, Brighton BN1
9QJ,~~UK\\[2mm]}
\end{center}

\hrule
\vskip 0.3cm
\begin{minipage}[h]{14.0cm}
\begin{center}
\small{\bf Abstract}\\[3mm]
\end{center}
Torsion appears due to fermions coupled to gravity and leads
to the strongest particle physics bounds on flat extra dimensions.
In this work, we consider torsion constraints in the case of a warped extra dimension
with  brane and bulk fermions.
From current data we obtain a 3$\sigma$ bound on the TeV--brane mass  
scale $\Lambda_\pi \geq 2.2~(10)$ TeV
for the AdS curvature $k=1 ~(0.01)$ in (reduced) Planck units.
If Dirac or light sterile neutrinos reside on the brane, the bound
increases to 17 (78) TeV.

\end{minipage}
\vskip 0.7cm
\hrule
\vskip 1cm

\section{Introduction.}

String theory requires the presence of extra dimensions. 
If some of these flat dimensions are large (submillimeter), the Standard Model (SM) 
hierarchy problem appears to be much milder \cite{Arkani-Hamed:1998rs}.
The extra dimensions can also be warped \cite{Randall:1999ee},
\cite{Randall:1999vf}, i.e. generate  a  non--zero  higher dimensional curvature.
In this case, the hierarchy problem can be solved given an appropriate brane
configuration in the higher dimensional space \cite{Randall:1999ee}.

Current experiments   require flat  extra dimensions to be 
of order micrometer or smaller
(see \cite{Rubakov:2001kp} for a recent review). These constraints come mainly from the phenomenology
of the Kaluza--Klein (KK) excitations of the graviton. Light KK gravitons are
emitted copiously in particle collisions and decays, and appear as missing energy and momentum
\cite{Mirabelli:1998rt},\cite{Giudice:1998ck}.  
Also, exchange of virtual KK gravitons generates corrections to the SM predictions
for collider observables  and leads to additional constraints \cite{Giudice:1998ck},
\cite{Agashe:1999qp}.
The fundamental higher--dimensional Planck scale is bounded by these experiments to be 
greater than about 1 TeV. 
Astrophysical considerations increase this scale to about 50-70 TeV for
two extra dimensions \cite{Cullen:1999hc}.
In the case of warped extra dimensions (the Randall--Sundrum model), 
the KK graviton modes are heavy but couple strongly to matter,
so they can be detected as massive spin 2 resonances.
Phenomenology of the Randall--Sundrum model  has been studied extensively 
\cite{Davoudiasl:1999jd},\cite{hewett} with the result that  current experiments constrain the mass scale on the visible brane to be  about 1 TeV or larger.

In these considerations, the gravitational connection $\Gamma_{\mu\nu}^\alpha$ was 
assumed to be symmetric.
Although this assumption leads to no inconsistencies, it is still an $assumption$.
An alternative approach is to make no {\it a priori} assumption and find
the gravitational connection through its equations of motion.
This is known as the first order or Palatini formalism. In the absence of fermions,
these two approaches are equivalent and lead to a symmetric connection.
However, if fermions are present, an antisymmetric piece or torsion is 
induced in the first order formalism \cite{Kibble:ba}.  Torsion (at least classically) is not a 
dynamical degree of freedom and can be eliminated from the action via its
equations of motion (for a recent review, see \cite{Buchbinder:rb}). 
The result is a four--fermion interaction suppressed in the four--dimensional case
by the Planck scale squared. The presence of torsion or, equivalently, this
contact interaction is in agreement with the standard relativity tests \cite{wheeler}
since its effect appears only in the presence of fermions and  does not directly affect
propagation of light. It is also worth mentioning that torsion is required in 
supergravity  \cite{Deser:1976eh}.

In four dimensions, the question whether torsion is present or not
is only of academical interest because its effects are enormously suppressed.
If extra space--time dimensions are present, the situation changes dramatically \cite{Chang:2000yw}. 
The torsion--induced contact interaction is suppressed only by the square of the
fundamental scale which could be of order TeV. For flat extra dimensions,
this enhancement has entirely different nature from that of the graviton
mediated interactions. It results from the large fermionic spin  density on the brane. 
This allows to obtain  the strongest
particle physics bounds on the higher dimensional fundamental scale.
A global fit to the LEP/SLD electroweak observables yields
a 3$\sigma$ bound \cite{Chang:2000yw}
\begin{equation}
M_S \geq 28 \;{\rm TeV}
\end{equation}
for $n=2$. It is worth emphasizing that this bound is obtained under a minimal set of assumptions.
In particular, it is based on the standard gravity action and  equations of motion for the
connection.
A more exotic possibility would be  to assume propagating (dynamical) torsion \cite{Belyaev:1997zv},
\cite{Mukhopadhyaya:2001fc}.

In the present work, we extend this analysis to the case of a warped extra
dimension. We find  that torsion effects provide a strong
bound on the fundamental scale, yet not as severe as for flat extra dimensions.
We  also generalize these results to the case of bulk fermions.

\section{Torsion in 4D (super)gravity.}

In this section, we will introduce our notation and provide basic facts about
torsion. 
We will follow the  conventions of Ref.\cite{VanNieuwenhuizen:ae}. The metric is $\eta_{\mu\nu}=
(++++)$ and the gamma matrices are hermitian.
The Lorentz--covariant derivative of the Majorana spinor $\chi$ 
is  defined by 
\begin{eqnarray}
&& D_\mu \chi=\left(\partial_\mu +{1\over 2} \sigma^{ab}\omega_{\mu a b }
\right)\chi \;.
\end{eqnarray}
with $\sigma^{ab}={1\over 4} (\gamma^a \gamma^b-\gamma^b \gamma^a)$.
The Lorentz connection $\omega_{\mu a b }$ is antisymmetric in the last two indices and
can  always be written as
a vierbein--dependent piece $\omega_{\mu a b }^0(e)$ and the contorsion $\kappa_{\mu a b }$:
\begin{equation}
\omega_{\mu a b }=\omega_{\mu a b }^0(e)+\kappa_{\mu a b }\;.
\label{decomposition}
\end{equation}
$\omega_{\mu a b }^0(e)$ can be obtained from the Einstein action in the absence of
fermions via the equations of motion for $\omega_{\mu a b }$ or, equivalently,
by imposing the ``tetrad postulate'' that the fully covariant derivative
(with the Christoffel connection)
 of the
veirbien vanish, 
\begin{equation}
{\mathcal D}^0_\mu e^m_\nu \equiv 
\partial_\mu e^m_\nu + \omega_\mu^{0~mn}(e)e_{n \nu}- \Gamma^{\alpha}_{\nu\mu} (g)
e^m_\alpha=0\;. 
\end{equation}
The contorsion tensor accounts for matter effects.
If the connection is considered an independent field and is found through its
equations of motion, the contorsion tensor does not vanish in the presence of fermions.
This is known as the Palatini or first-order formalism. 
It is advantageous in that it requires no {\it a priori} assumptions about the 
properties of the connection. 

On the other hand,
one may $assume$ that the connection is always symmetric. This possibility is 
self-consistent and is motivated by the equivalence principle in its {\it very strong}
form, i.e. that all gravity effects (up to higher order corrections) 
can be eliminated locally by an 
appropriate coordinate transformation \cite{wheeler}. Since contorsion (or torsion) 
is a tensor, it cannot be eliminated in this way. 
However, the  assumption  of local removability of ``all'' gravitational effects
may be too strong.
Indeed, the presence of torsion does not directly affect propagation of test 
particles and is in perfect agreement with the standard general relativity tests.
To detect the presence of torsion would require a detailed investigation
of the spin--gravitational effects with fermionic test particles.
Thus, there is no compelling reason to assume that the connection stays symmetric
in the presence of fermions.
Also, we note that it is the first order formalism that leads to the standard
description of supergravity, so torsion is present in locally
supersymmetric theories \cite{Deser:1976eh}.

Let us now consider how torsion arises in four dimensional gravity  and supergravity.

{\em {\bf 1.~4D Gravity.}} The Lagrangian of gravity coupled to a Majorana fermion 
$\chi$ is given by 
\begin{equation}
{\mathcal L}= -{1\over 2 \kappa^2} e R 
-{1\over 2}e \bar \chi {\skew{11}\not{D}} \chi \;,
\end{equation}
where $\kappa = \sqrt{8\pi}/M_{\rm Pl}$, $e~$=$~$det$~e^m_\mu$, and $R=e^{n\mu} e^{m\nu}
R_{\mu\nu m n}$ with the curvature tensor
\begin{equation}
R_{\mu\nu}^{~~\;mn}=\partial_\mu \omega_\nu^{~mn}+\omega_\mu^{~mc}\omega_{\nu c}^{~~n}
-(\mu \leftrightarrow \nu)\;.
\end{equation}
Using the decomposition (\ref{decomposition}), we have
\begin{equation}
R=R(e)+ e_n^\mu e_m^\nu D^0_{[\mu}\kappa_{\nu ]}^{~mn}- \kappa_{\mu\rho\nu} \kappa^{\nu\rho\mu}
+(\kappa_\nu^{~\nu\rho})^2 \;,
\label{r}
\end{equation}
where $[\mu\nu]\equiv \mu\nu-\nu\mu$ and $R(e)$ is built on $\omega_{\mu a b }^0(e)$.
In the second term of this equation, the Lorentz--covariant
derivative $D^0_\mu$ built on $\omega_{\mu a b }^0(e)$  can be replaced by 
the fully covariant derivative ${\mathcal D^0_\mu}$ built on $\omega_{\mu a b }^0(e)$
and the Christoffel symbols $\Gamma^{\alpha}_{\nu\mu} (g)$. Using 
${\mathcal D}^0_\mu e^m_\nu =0$, it is easy to see that 
this term is a total divergence. Therefore, the Einstein action has an algebraic
 dependence on the contorsion tensor. It is easy to see that
the contorsion derivatives do not appear anywhere and thus contorsion is a non-propagating
field which can be eliminated algebraically. 

The Einstein action has a quadratic dependence on the contorsion while 
the fermion action depends on it linearly.
Thus, varying the action with respect to $\kappa^{\nu\rho\mu}$ 
gives{\footnote{All $\epsilon$--tensors are assumed to take on the values $\pm 1,0$.
Thus, lowering the world indices of the $\epsilon$--tensor is accompanied by
dividing it by det$~g_{\mu\nu}$ and $\epsilon^{\mu\nu\rho\sigma }=
({\rm det}~ e_m^\mu)^{-1} e_m^\mu e_n^\nu e_k^\rho e_l^\sigma  
\epsilon^{mnkl}$.}}
\begin{equation}
\kappa_{\mu\nu\rho}- \kappa_{\rho\nu\mu}=
 - {e \kappa^2 \over 4}
\epsilon_{\mu\nu\rho\sigma} \bar \chi \gamma^5 \gamma^\sigma \chi \;.
\end{equation}
This equation can be solved for contorsion:
\begin{equation}
\kappa_{\mu\nu\rho}=
- {e \kappa^2 \over 8} \epsilon_{\mu\nu\rho\sigma}
\bar \chi \gamma^5 \gamma^\sigma \chi \;.
\label{cont}
\end{equation}
The torsion tensor is defined as the antisymmetric part of the connection $\Gamma^{\alpha}
_{\mu\nu}$ which is related to the spin--connection via
$\partial_\mu e^m_\nu + \omega_\mu^{~mn}e_{n \nu}- \Gamma^{\alpha}_{\nu\mu} 
e^m_\alpha=0$,
\begin{equation}
S^\alpha_{\mu\nu}= {1\over 2} \Bigl(\Gamma^{\alpha}_{\mu\nu}-\Gamma^{\alpha}_{\nu\mu}
\Bigr)=
{1\over 2}\Bigl(-\kappa_{\mu~\nu}^{~\alpha}+\kappa_{\nu~\mu}^{~\alpha} \Bigr)\;. 
\end{equation}
Clearly, the torsion tensor is proportional to $D_\mu e^m_\nu -D_\nu e^m_\mu $.
From Eq.\ref{cont} we obtain 
\begin{equation}
S_{\mu\nu\alpha}= - {e \kappa^2 \over 8} \epsilon_{\mu\nu\alpha\sigma}
\bar \chi \gamma^5 \gamma^\sigma \chi \;.
\end{equation}
We see that (con)torsion vanishes outside matter distribution and is completely
antisymmetric if only spin 1/2 fermions are present. 

Using Eqs.\ref{r} and \ref{cont}, the contorsion tensor can be eliminated from the action.
This results in the four-fermion axial interaction
\begin{equation}
\Delta {\mathcal L}= {3e\kappa^2 \over 64} \Bigl(\bar \chi \gamma^5 \gamma^\sigma \chi
\Bigr)^2 \;.
\label{interaction}
\end{equation}
Interaction of this form is specific to torsion and cannot be induced by
the graviton exchange. It is repulsive for aligned spins. Indeed, in the 
non-relativistic limit the corresponding Hamiltonian reads
\begin{equation}
\Delta {\mathcal H}=-{3e\kappa^2 \over 64} \Bigl(\bar \chi \gamma^5 \gamma^\sigma \chi
\Bigr)^2 \longrightarrow {3e\kappa^2 \over 64}
(u^\dagger \stackrel{\rightarrow}{\sigma} u)^2 \;,
\end{equation}
where $u$ is a two-component non-relativistic spinor.

In the case of many Majorana fields, Eq.\ref{interaction} generalizes to 
\begin{equation}
\Delta {\mathcal L}= {3e\kappa^2 \over 64} \Bigl( \sum_i \bar \chi_i \gamma^5 \gamma^\sigma \chi_i \Bigr)^2 \;.
\end{equation} 
It is useful to convert this interaction into that for the Dirac fermions.
Expressing $\Psi= P_L \chi_1 +P_R \chi_2$ with $P_{L,R}$ being the left and right 
projectors and using $\bar \chi \gamma^\mu \chi=0$, we have{\footnote{In the notation of Ref.\cite{Chang:2000yw}, the coefficient of this contact interaction would be $3e \kappa^2/32$
since $\kappa^2$ defined here is twice as small. }}
\begin{equation}
\Delta {\mathcal L}= {3e\kappa^2 \over 16} \Bigl( \sum_i \bar \Psi_i \gamma^5 \gamma^\sigma \Psi_i \Bigr)^2 \;,
\label{diracinteraction}
\end{equation}
where the sum now runs over Dirac fermions.
The most peculiar feature of this interaction is that it is completely universal
for all of the spin 1/2 fermions. In the case of the Standard Model, it possesses
the maximal possible symmetry $U(45)$ acting on the 45 Weyl spinors \cite{Buchmuller:1997hn}.
This universality stems from the fact that torsion couples
to the spin density. In contrast, the four--fermion interactions induced
by the graviton exchange would depend on the energy and masses of the fermions
since the graviton couples to the energy--momentum tensor.
The interaction (\ref{diracinteraction}) is a truly gravitational effect
which can hardly be ``counterfeited'' by other physics.

Whereas in the case of gravity the presence of torsion is logical
yet optional, supergravity theory requires torsion \cite{Deser:1976eh}. 
Let us now
consider how the above equations get modified in supergravity.

{\em {\bf 2.~4D Supergravity.}}
The Lagrangian of supergravity coupled to a chiral supermultiplet  
in the first order formulation is given by \cite{Ferrara:1976kg}
\begin{eqnarray} 
{\mathcal L}&=& -{1\over 2 \kappa^2} e R -{1\over 2}\epsilon^{\mu\nu\rho\sigma}
\bar \Psi_\mu \gamma_5 \gamma_\nu  D_\rho \Psi_\sigma
-{1\over 2}e \bar \chi {\skew{11}\not{D}} \chi - 
{1\over 16}e \kappa^2 \Bigl(\bar \chi \gamma^5 \gamma^\sigma \chi
\Bigr)^2  \nonumber\\
&+& {\rm connection\;independent\;terms }\;.
\label{super}
\end{eqnarray}
Here $\Psi_\nu$ is the gravitino field and 
\begin{equation}
 D_\mu \Psi_\nu=\left(\partial_\mu +{1\over 2} \sigma^{ab}\omega_{\mu a b }
\right) \Psi_\nu \;.
\end{equation}
Even though this derivative 
 is  not covariant in the world indices, its $curl$ is, so the above action
is indeed invariant under coordinate transformations.
The supergravity multiplet, the spin--connection, and the matter multiplet are
considered to be independent. Each of them has its own supersymmetry
transformation (see e.g.\cite{Ferrara:1976kg}). 
An interesting feature of the above Lagrangian is that supersymmetry
requires the presence of the four--fermion axial interaction which appears
independently from the one induced by torsion.

Variation of the action with respect to contorsion is done the same way discussed above
except now there is an additional contribution from the gravitino field.
The field equation for contorsion is
\begin{equation}
\kappa_{\mu\nu\rho}- \kappa_{\rho\nu\mu}=
{\kappa^2 \over 2} \bar \Psi_\mu \gamma_\nu \Psi_\rho - {e \kappa^2 \over 4}
\epsilon_{\mu\nu\rho\sigma} \bar \chi \gamma^5 \gamma^\sigma \chi \;.
\end{equation}
Solving for contorsion, we get
\begin{equation}
\kappa_{\mu\nu\rho}={\kappa^2 \over 4}\left( \bar\Psi_\mu \gamma_\nu \Psi_\rho+
\bar\Psi_\nu \gamma_\mu \Psi_\rho -\bar\Psi_\mu \gamma_\rho \Psi_\nu \right)
- {e \kappa^2 \over 8} \epsilon_{\mu\nu\rho\sigma}
\bar \chi \gamma^5 \gamma^\sigma \chi \;.
\label{cont1}
\end{equation}
Note that in the presence of spin 3/2 particles the contorsion and torsion
tensors are no longer completely antisymmetric.
Eliminating contorsion from the action, we get the same four--fermion interaction
as in the case of gravity plus additional terms involving the gravitino.
In addition to this torsion--induced interaction, we have the original four--fermion
term in Eq.\ref{super}. The final result is
\begin{equation}
\Delta {\mathcal L}=- {e\kappa^2 \over 64} \Bigl(\bar \chi \gamma^5 \gamma^\sigma \chi
\Bigr)^2 + {\rm gravitino\;terms}\;.
\label{interaction1}
\end{equation}
It is remarkable that, compared to the previous case, not only the numerical coefficient
has changed but also the interaction has flipped its sign.
That means that if we start with supergravity, break it spontaneously, and
integrate out the superpartners, the residual four-fermion interaction
will be different from that of  gravity.
Another interesting feature of supergravity is that if we start with a vector
supermultiplet coupled to supergravity, the resulting four--gaugino interaction
turns out to be  $\Delta {\mathcal L}= 3e\kappa^2/16 ~(\bar \lambda \lambda)^2$
instead of (\ref{interaction1}).

The torsion--induced interaction in four dimensions is suppressed by $M_{\rm Pl}^2$
and thus is undetectable. However, if there are additional large space--time
dimensions, the situation changes drastically and the torsion effects become
not only visible but even dominant \cite{Chang:2000yw}.
In the next section, we consider torsion effects in the case of a warped extra dimension,
i.e. the Randall-Sundrum model.

\section{Torsion in the Randall-Sundrum model.}

The Randall--Sundrum setup provides an attractive way to generate a hierarchy
between the electroweak and the Planck scales \cite{Randall:1999ee}. The basic idea is to start
with a five--dimensional anti-de Sitter (AdS) space--time with two 3-branes.
With a special choice of the vacuum energies in the bulk and on the branes,
this configuration can be made stable. 
The distance between the branes is determined by a VEV of the radion field \cite{Goldberger:1999uk}.
Then, the geometrical warp factor
representing the overlap of the graviton wave function with the observable
brane
is responsible  for the apparent weakness of gravity.

The AdS metric
\begin{equation}
ds^2=e^{-2 k \vert y \vert}\eta_{\mu\nu}dx^\mu dx^\nu + dy^2
\end{equation}
is induced by the gravitational fields of the branes.
Here $y$ ~($-\pi r < y \leq \pi r$) parametrizes the orbifold extra dimension of
radius $r$ and  $y \equiv -y$.  This metric is a solution to the Einstein
equations if{\footnote{We use the notation of Ref.\cite{Gherghetta:2000qt}.}}
\begin{eqnarray}
&& S_{\rm bulk}= -\int d^5 x ~\hat e \left(  {1\over 2} M_5^3~ R +\Lambda \right) \;, \nonumber\\
&& S_{\rm brane}= -\int d^5 x ~e \Bigl[\delta (y) \left( \Lambda_1 + {\cal L}_1 \right)
+ \delta (y-\pi r) \left( \Lambda_2 + {\cal L}_2 \right)
 \Bigr]\;,
\end{eqnarray}
with 
\begin{eqnarray}
&& \Lambda= -6 M_5^3 k^2 \;, \nonumber\\
&& \Lambda_1 =- \Lambda_2= -{\Lambda \over k}\;.
\end{eqnarray}
Here $1/k$ is the AdS curvature radius and $M_5$ is the five--dimensional Planck mass.
They are related to the four--dimensional reduced Planck mass $\tilde M_{\rm Pl}=M_{\rm Pl}/
\sqrt{8\pi}$ via
\begin{equation}
\tilde M_{\rm Pl}^2= {M_5^3 \over k}(1- e^{-2 \pi k r})\;.
\end{equation}
We note that the gravitational constant $\kappa$ introduced in the previous section
is given by $\kappa= 1/ \tilde M_{\rm Pl}$. 
With the above metric it can be easily shown that the natural mass scale at $y=0$
is $M_{\rm Pl}$ while that at $y=\pi r$ is $M_{\rm Pl}~e^{-\pi k r}$.
Thus, with $kr\simeq 12$, we obtain a TeV scale at $y=\pi r$ (the visible brane). 
The mass hierarchy in this case has  a  geometrical origin -- it appears owing
to the AdS metric in the five--dimensional space.

Let us now consider how torsion arises in this setup. We will discuss
separately torsion effects induced by brane fermions and fermions proparating
in the bulk.

{\bf i. Brane fermions.} The visible brane contains 
Majorana fermions (which can be converted into Dirac ones)
with the kinetic terms
\begin{equation}
S_{\rm brane \; ferm.}= \int d^5 x~ \delta (y- \pi r) \left[ 
-{1\over 2}e \sum_i \bar \chi_i {\skew{11}\not{D}} \chi_i \right] \;.
\label{kinetic}
\end{equation}
These are the Standard Model fermions localized on the brane.
To find the equations of motion for the connection, we vary the action with respect to 
contorsion noting  that Eq.\ref{r} is valid for an arbitrary number of space--time 
 dimensions.
The result is 
\begin{equation}
\kappa_{\mu\nu\rho}=
- {e~ \delta(y-\pi r) \over 8 M_5^3} \epsilon_{\mu\nu\rho\sigma} \sum_i
\bar \chi_i \gamma^5 \gamma^\sigma \chi_i \;.
\end{equation}
Here we have used $\hat e=e$ in the AdS background.
The other torsion components are zero.
Eliminating non--propagating
contorsion from the action, we obtain the following
axial contact interaction on the visible brane:
\begin{equation}
\Delta {\mathcal L}= {3e~\delta(0) \over 64 M_5^3} \Bigl( \sum_i \bar \chi_i \gamma^5 \gamma^\sigma \chi_i \Bigr)^2 \;.
\end{equation}
The arising delta--function singularity is due to the implicit assumption that the brane
is infinitely thin. (Con)torsion is proportional to the fermionic spin--density, which
from the five--dimensional point of view becomes infinite on the brane. 
In practice, the delta--function should be regularized to account for a finite brane width:
\begin{equation}
\delta(0) \rightarrow {1\over 2 \pi}\int_{-M_5}^{M_5} dk= {M_5 \over \pi} \;,
\label{delta}
\end{equation}
where we have taken the five--dimensional Planck mass as a natural cut--off
so that the brane width is of order $M_5^{-1}$.

The metric on the visible brane is given by $e^{-2k\pi r}\eta_{\mu \nu}$. Thus, 
the fermion kinetic terms in Eq.\ref{kinetic} are not canonically normalized:
\begin{equation}
-{1\over 2}e \bar \chi {\skew{11}\not{D}} \chi =
-{1\over 2} e^{-2k\pi r}\eta^{\mu\nu} ~\bar \chi \gamma_\mu D_\nu \chi\;.
\end{equation}
Rescaling $\chi \rightarrow \chi ~ e^{k\pi r} $ and using Eq.\ref{delta}, 
we obtain the following
four--fermion interaction for the properly normalized fields:
\begin{equation}
\Delta {\mathcal L}= {3 \over 64 \pi ( e^{-k \pi r}M_5)^2} \Bigl( \sum_i \bar \chi_i \gamma^5 \gamma^\sigma \chi_i \Bigr)^2 ={3 \over 16 \pi ( e^{-k \pi r}M_5)^2} \Bigl( \sum_i \bar \Psi_i \gamma^5 \gamma^\sigma \Psi_i \Bigr)^2 \;,
\label{torsioninteraction}
\end{equation}
where $\Psi_i$ are the Dirac brane fermions. 
As we see, the torsion--induced interaction is suppressed only by the TeV scale, unlike
in the 4D case. 
The reason is that $e^{-k \pi r}M_5$ can be viewed as the fundamental scale \cite{Randall:1999ee}.
Also, 
 the large five--dimensional spin--density on the brane plays an important role.
If we allow some of the fermions to propagate in the bulk, the
torsion--induced interaction of their zero modes will be suppressed by the ``volume'' factor
$k r \sim {\cal O}(10)$. This differs from the case  of flat extra dimensions 
in which case
large spin--density on the brane was the only reason for enhancement of the torsion--induced
interaction.

We note that the Kaluza-Klein graviton exchange also generates four--fermion interactions.
These are however further suppressed by $E^2/( e^{-k \pi r}M_5)^2$ with $E$ being the
typical energy of the process since the graviton couples to the energy--momentum
tensor \cite{Giudice:1998ck}. Furthermore, the graviton exchange can induce neither  axial nor
universal contact interaction.

The  axial interaction (\ref{torsioninteraction}) induces significant vertex 
corrections to the
Z couplings to fermions and is strongly constrained by the electroweak precision data.
The crucial point is that it is $universal$ for all of the fermions and contains
no variable parameters apart from the mass scale.
The global fit to the LEP/SLD observables was performed in Ref.\cite{Chang:2000yw}. It was found
that this interaction is excluded at the 2$\sigma$ level by classical statistical analysis 
and is allowed only at the 3$\sigma$ level. This occurs mainly because (\ref{torsioninteraction})
increases the $Z\rightarrow \nu \bar \nu$ width whereas its experimental value is about
2$\sigma$ smaller than the Standard Model prediction{\footnote{On possible relevance of
supersymmetry to this deviation see Ref.\cite{Lebedev:2001ez}.}}.
The major effect comes from the vertex correction with the top quark in the loop.
The corresponding shift in the Z--couplings is then
$\delta h_L =-\delta h_R=3 N_c m_t^2/(64 \pi^3 M_*^2) \ln M_*^2/m_t^2$,
where for brevity we denoted $M_* \equiv e^{-k \pi r}M_5$.
The resulting
3$\sigma$ bound is $e^{-k \pi r}M_5 \geq 2.2$ TeV, or in the notation of Ref.\cite{Davoudiasl:1999jd},
\begin{equation}
\Lambda_\pi \left( {k\over \tilde M_{\rm Pl} }\right)^{1/3} \geq 2.2 \;{\rm TeV}\;,
\label{bound}
\end{equation}
where $\Lambda_\pi =\tilde M_{\rm Pl~}e^{-k \pi r}$ and ${\tilde M_{\rm Pl}}^2\simeq M_5^3/k$.
The Randall--Sundrum solution is valid for $k \leq \tilde M_{\rm Pl}$. If 
$k \simeq \tilde M_{\rm Pl}$, we have $\Lambda_\pi \geq 2.2$ TeV, while for 
$k \simeq 0.01 \tilde M_{\rm Pl}$ the constraint becomes significantly stronger:
$\Lambda_\pi \geq 10$ TeV.

An additional (tree-level) constraint on the interaction (\ref{torsioninteraction}) 
can be obtained from the OPAL measurements of the differential
cross sections $e^+e^- \rightarrow f \bar f$ \cite{Alexander:1996rg}. These imply
\begin{equation}
\Lambda_\pi \left( {k\over \tilde M_{\rm Pl} }\right)^{1/3} \geq 0.8 \;{\rm TeV}
\end{equation}
at the 95\% confidence level. Constraints from DIS data, Drell--Yan production, etc. are
weaker \cite{Chang:2000yw}.

The 3$\sigma$ bound (\ref{bound}) is quite significant. For instance, with $k \sim \tilde M_{\rm Pl}$, it  
is stronger than the 95\% C.L. bound obtained from collider data in Ref.\cite{Davoudiasl:1999jd}
by about a factor of five.  Yet, torsion constraints on the fundamental scale
are even more severe in
the case of flat extra dimensions, e.g. $M_S \geq 28 \;{\rm TeV}$ for $n=2$.
The reason is that  for flat extra dimensions the spin--density on the brane is more singular
because of the large hierarchy between the compactification scale and  the fundamental scale 
(apart from the fact that here we constrain the ``reduced'' fundamental scale).
In the Randall--Sundrum scenario, this hierarchy is only one-two orders of magnitude,
so the effect of $\delta(0)$ is not as significant.  

Finally, there is a strong astrophysical bound on the interaction (\ref{torsioninteraction}).
If the Dirac or light  ($m_{\nu_s} \ll 50$ MeV) sterile neutrinos live on the brane,
torsion induces 
\begin{equation}
\Delta {\mathcal L} =
{6 \over 16 \pi ( e^{-k \pi r}M_5)^2}  ~\bar q \gamma^5 \gamma^\sigma q ~
 \bar \nu_R  \gamma^5 \gamma_\sigma \nu_R \;.
\label{SN}
\end{equation}
A quark--neutrino contact interaction  of this type provides a new channel of energy drain
during neutron star collapse and is severely constrained by the supernova data \cite{Grifols:1997iy}.
The corresponding bound from SN 1987A is
\begin{equation}
\Lambda_\pi \left( {k\over \tilde M_{\rm Pl} }\right)^{1/3} \geq 17 \;{\rm TeV}\;.
\end{equation}
As we will see below, this bound  relaxes if we allow neutrinos to propagate
in the bulk. 

We note that the only source of uncertainty
in the above (classical) calculations is the brane width or, equivalently, regularization of the  
delta function. Our bounds are inversely proportional to the square root of the   
brane width and, thus, are not very sensitive to this source of uncertainty. 
If the brane width exceeds $M_5^{-1}$ considerably, the common $\delta$-function 
description of the brane breaks down as the brane width  would be comparable 
to the compactification radius.  
In addition, there are, of course, ever--present quantum corrections. 
For instance, the fermion--torsion coupling $S_\mu \bar \Psi \gamma^5 \gamma^\mu \Psi$,
where $S_\mu \propto  \epsilon_{\mu\nu\rho\sigma} \kappa^{\nu\rho\sigma}$,
induces the torsion kinetic terms $\partial_{[ \mu} S_{\nu ]}$ via the fermion loop.
Thus, torsion becomes dynamical and can propagate along the brane. Our approximation
corresponds to shrinking the torsion propagator into a point. Since the torsion mass
coming from the scalar curvature
is of order the cut--off scale, this approximation is  very good at low  energies.

{\bf ii. Bulk fermions.}
Let us now consider torsion effects due to bulk fermions. 
A straightforward modification of the kinetic terms (\ref{kinetic})
$\gamma^\mu D_\mu \rightarrow \gamma^\mu D_\mu +\gamma^5 D_5 $
 leads to the following
equations of motion for contorsion:
\begin{eqnarray}
&& \kappa_{\mu\nu\rho}=
- {e~  \over 4 M_5^3} \epsilon_{\mu\nu\rho\sigma} \sum_i
\bar \Psi_i \gamma^5 \gamma^\sigma \Psi_i \;, \nonumber\\
&& \kappa_{5\nu\rho}=
- {1  \over 2 M_5^3}  \sum_i
\bar \Psi_i \gamma_5 \sigma_{\nu\rho} \Psi_i \;, \nonumber\\
&& \kappa_{55\rho}=0\;,
\end{eqnarray}
where  $\Psi_i$ are the five--dimensional Dirac fermions. Note that $\kappa_{5\nu\rho}$
is completely antisymmetric since $\gamma_5 \sigma_{\nu\rho}=
\gamma_{[5} \gamma_{\nu} \gamma_{\rho ]}/12$, so the other components of the contorsion tensor
with two four--dimensional indices are given by permutations. 
It is interesting to note that new (tensor) structures have appeared compared to
the 4D fermion case.

To obtain chiral fermions in four dimensions, the Dirac spinor is taken to obey the
orbifold boundary condition $\Psi (-y)=\pm \gamma_5 \Psi (y)$ (see e.g. \cite{Flachi:2001bj}). Since $\bar \Psi \gamma_5 
\sigma_{\nu\rho} \Psi $ is odd under $y \rightarrow -y$, $\kappa_{5\nu\rho}$ vanishes
at the orbifold fixed points.
Thus, the torsion--induced  interaction for the SM fermions
 has the same structure as before, i.e. an axial vector squared.

The observed fermions correspond to the massless (zero) modes of the 5D fermions.
The properly normalized zero modes are given by \cite{Gherghetta:2000qt}
\begin{equation}
\Psi (x,y)= {e^{(1/2-c)k\vert y \vert} \over \sqrt{2\pi r} N_0} \Psi (x)\;,
\end{equation}
with $N_0$ being the normalization factor,
\begin{equation}
N_0^2={e^{2\pi k r (1/2-c)}-1 \over 2\pi k r (1/2 -c)}\;.
\end{equation}
Here the constant $c$ indicates the localization of the zero mode
in the AdS space:
for $c < 1/2$ the zero mode peaks on the TeV--brane, 
for $c > 1/2$ it is localized towards the Planck ($y=0$) brane,
and for $c=1/2$ it is constant in the $y$--direction.

For clarity, we assume $c=1/2$ in which case $N_0=1$ and 
$\Psi (x,y)={1\over \sqrt{2\pi r}} \Psi (x)$.
Then, the torsion--induced interaction is given by
\begin{equation}
\Delta {\mathcal L} ={3 \over 32 \pi ( e^{-k \pi r}M_5)^2 rM_5}  
\Bigl( \sum_i \bar \Psi_i \gamma^5 \gamma^\sigma \Psi_i \Bigr)^2 \;,
\end{equation}
Compared to the previous case, the delta function $\delta(0)=M_5/\pi$ 
got replaced by the volume factor $(2\pi r)^{-1}$. This results in an order
of magnitude suppression. Since $M_5 r \simeq 12 M_5/k = 12 (\tilde M_{\rm Pl} 
/k)^{2/3}$ and $( e^{-k \pi r}M_5)^2=\Lambda_\pi^2 (\tilde M_{\rm Pl} 
/k)^{-2/3}$, we have
\begin{equation}
\Delta {\mathcal L} \simeq {1 \over 128  \pi \Lambda_\pi^2}  
\Bigl( \sum_i \bar \Psi_i \gamma^5 \gamma^\sigma \Psi_i \Bigr)^2 \;.
\label{bulk}
\end{equation}
The resulting bounds on $\Lambda_\pi$ are about a factor of 5 weaker than
in the case of localized fermions (with $k \simeq \tilde M_{\rm Pl}$). 
As mentioned above, this is to be contrasted with the case of flat extra dimensions:
if the SM fermions are allowed to propagate in the extra dimensions, no appreciable
torsion constraint can be obtained.

Finally, the interaction of two brane and two bulk fermions is the same as if
they all were bulk fermions of different species and  is given by Eq.\ref{bulk}.
This possibility is interesting since a coupling between a neutral bulk fermion and
the brane neutrino and Higgs induces a naturally small Dirac neutrino mass 
for flat extra dimensions
\cite{Arkani-Hamed:1998vp}{\footnote{One should keep in mind that
this scenario is strongly constrained by the supernova data \cite{Barbieri:2000mg}.}}.
In this case torsion effects require $M_S \geq 28$ TeV (ADD, $n=2$) and $\Lambda_\pi \geq 3$ TeV (RS).

\section{Discussion and conclusions.}

We have considered the effects of torsion in the Randall--Sundrum model with two branes.
Our analysis was based on the first order formalism. That is, instead of making an {\it a priori}
assumption that the gravitational connection is symmetric, we determined the properties
of the connection via its equations of motion. This resulted in a universal axial contact interaction
suppressed by a TeV scale. Assuming that the SM fermions are confined to the brane, 
current LEP/SLD electroweak data constrain the mass scale of the visible brane $\Lambda_\pi$:
\begin{equation}
\Lambda_\pi \left( {k\over \tilde M_{\rm Pl} }\right)^{1/3} \geq 2.2 \;{\rm TeV}\;.
\end{equation}
If Dirac or light sterile neutrinos also reside on the brane, this bound increases to 
17 TeV from the supernova SN1987 observations.
The bounds relax by roughly a factor of 5 if the fermions are allowed to propagate 
in the bulk.

It would be interesting to extend this analysis to the supersymmetric Randall--Sundrum
model. 
Important  steps for the case of pure supergravity have been made in this direction in Ref.\cite{Altendorfer:2000rr} (see also \cite{Gherghetta:2000qt}),
however the four--fermion terms were neglected.
It is curious that if the sign of the contact interaction flips upon supersymmetrizing,
as it does in four dimensions (Eq.\ref{interaction1}), this interaction 
 will be preferred by the electroweak data
and will help rectify the deviation in the invisible Z width. The other constraints will however
persist.

{\bf Acknowledgements.} This research was supported by PPARC.

\end{document}